\documentclass[aps, prl, 10 pt, notitlepage, twocolumn, letterpaper, preprintnumbers, longbibliography, floatfix]{revtex4}

\pdfoutput=1
\usepackage{tikz}
\usepackage[compat=1.0.0]{tikz-feynman}

\usepackage[utf8]{inputenc}
\usepackage[normalem]{ulem}
\usepackage{graphicx}% Include figure files
\usepackage{dcolumn}% Align table columns on decimal point
\usepackage[colorlinks=true,allcolors=purple]{hyperref}
\usepackage{url}
\usepackage{enumerate}
\usepackage{slashed,multirow,relsize,soul,feynmp-auto,tikz}
\usepackage{color}
\usepackage{mathrsfs} % pretty maths
\usepackage{amsmath}
\usepackage{cancel}
\usepackage{bbold}
 \usepackage{mathrsfs}
\usepackage{braket}
\usepackage{physics}
\usepackage{multirow}
\usepackage[capitalize]{cleveref}
\usepackage{xspace}

\hypersetup{colorlinks,citecolor= nicered,linkcolor= nicered}
\definecolor{nicered}{rgb}{0.7,0.1,0.1}
\definecolor{nicegreen}{rgb}{0.1,0.5,0.1}
\begin{document}
\def\Carleton{Department of Physics, Carleton University, Ottawa, Ontario K1S 5B6, Canada}

\title{String Decomposition and Gravitational Waves in High-quality Axion Gauge Theories}

\author{Camilla Mupo}
\email{CamillaMupo@cmail.carleton.ca}
\author{Yue Zhang}
\email{yzhang@physics.carleton.ca}
\affiliation{\Carleton}
\date{\today}
\begin{abstract}
The QCD axion can successfully solve the strong CP problem under the condition that the Peccei-Quinn symmetry is respected to extremely high standard. We explore a class of gauge theories that accommodate a high-quality axion, known as the Barr-Seckel models, paying special attention to the cosmology of topological defects. In models with domain wall number equal to unity, we show that axion strings with winding number larger than one can always be decomposed into a number of unit-winding axion strings and pure gauge strings. This mechanism enables the axion string-wall network to be destroyed timely to render a viable cosmology. The subsequent decay of gauge strings into gravitational waves generically produces a double-plateau in frequency space allowing the mechanism to be tested by upcoming experiments.
\end{abstract}
\maketitle

The strong CP problem is a naturalness problem of the Standard Model (SM) raised by the non-observation of neutron electric dipole moment (EDM). 
The introduction of QCD axion~\cite{Peccei:1977hh, Peccei:1977ur, Weinberg:1977ma, Wilczek:1977pj} provides a dynamical solution and rich opportunities for experimental searches~\cite{Marsh:2015xka, Irastorza:2018dyq, Sikivie:2020zpn, Adams:2022pbo}.
Of central importance to the solution is a $U(1)$ global symmetry.
Essentially, the puzzle about smallness of the $\bar\theta$ parameter is transmuted into how the Peccei-Quinn (PQ) symmetry can be upheld to a very high quality~\cite{Holman:1992us, Kamionkowski:1992mf, Barr:1992qq, Ghigna:1992iv}.
This is often made possible by resorting to additional symmetries, and gauge theories stand out for their power in restricting the form of Planck slop operators.

\begin{figure}[h]
  \includegraphics[width=1\columnwidth]{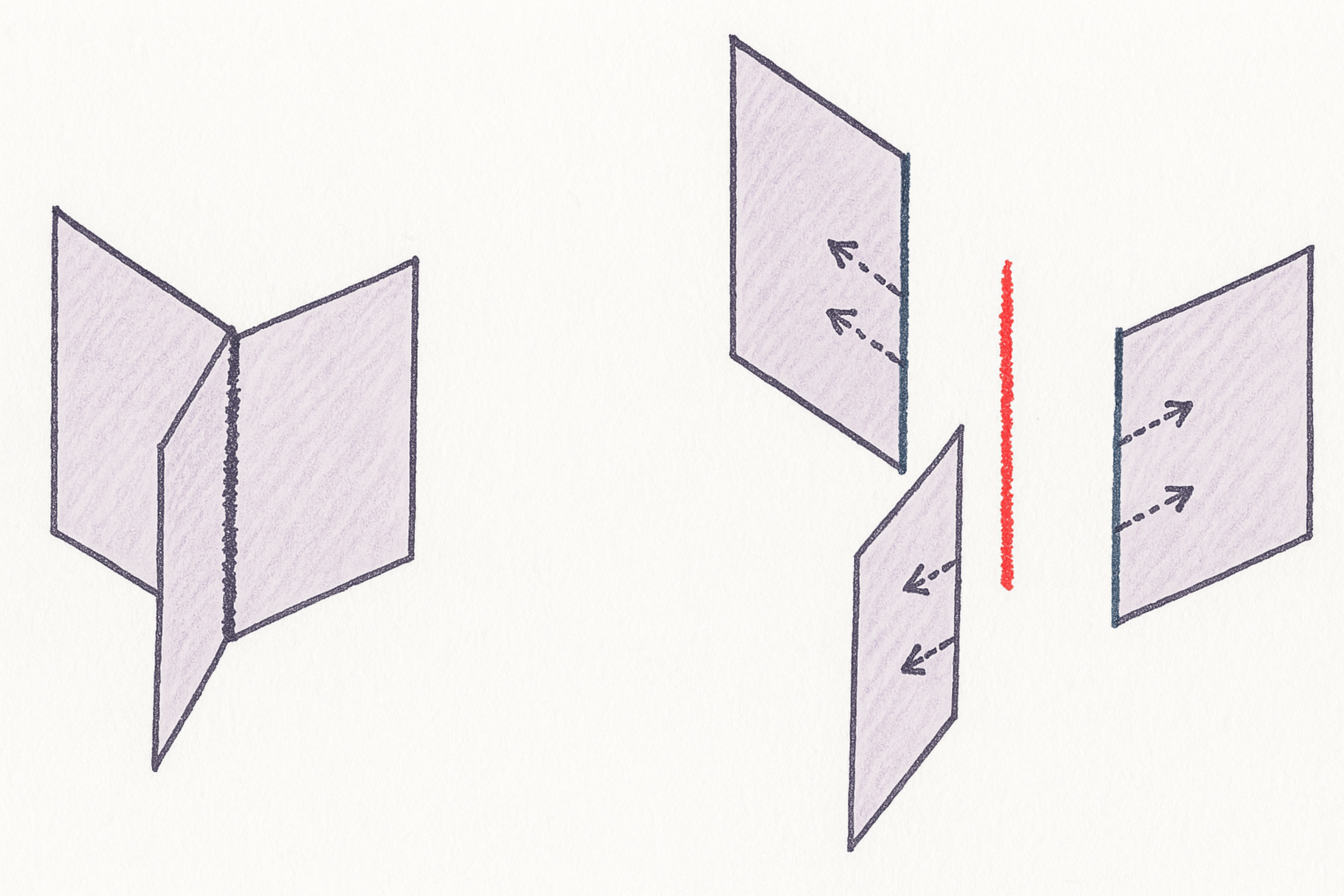}
  \caption{Schematic plot of axion string decomposition triggered by domain walls. As a concrete case, it corresponds to the first row of table~\ref{tab1}, where a string has axion winding number $w_a=3$ in a Barr-Seckel model with $p=3, q=7$. {\it Left panel}: when the QCD potential turns on, the string is originally the junction of three axion domain walls. {\it Right panel}: around time $t_2$, the string decays following the rule presented in Eq.~\eqref{eq:decomposition}, into three unit-winding axion strings, each serving as the edge of one domain wall, plus one pure gauge string (red). After the collapsing away of the axion-string-wall system, the gauge string is left over to produce gravitational waves.} \label{fig:picture}
\end{figure}

In this letter, we look into a class of gauge theories that can accommodate a high quality axion, originally written down by Barr and Seckel~\cite{Barr:1992qq}.
Their trick is to introduce a $U(1)$ gauge symmetry spontaneously broken by two scalar fields with distinctive charges which also source the masses of heavy quarks.
If the symmetry breaking occurs after inflation, an important consequence is the production of gauge and axion strings. 
We revisit the fate of these topological defects in the early universe, including their evolution into one another, and the interplay with domain walls.
We point out that a class of the Barr-Seckel models, that previously were thought to suffer from cosmologically long-lived domain wall problem, are actually viable. 
This is based on a string decomposition rule that will be presented below and illustrated by Fig.~\ref{fig:picture}.
Our investigation reveals a generic prediction in the isotropic gravitational wave (GW) background from gauge string radiations that serves as a smoking gun signature of these theories.

Our starting point is the interacting Lagrangian
\begin{equation}\label{eq:Yuk}
\mathcal{L} = \sum_{i=1}^p y \Bar{Q}^i_{\rm L} Q^i_{\rm R} S +\sum_{j=1}^q y' \Bar{Q}'^j_{\rm L} Q'^j_{\rm R} T + {\rm h.c.} \ ,
\end{equation}
where $S$, $T$ are SM gauge singlet scalars that carry charges $q_s$, $q_t$ under a new $U(1)$ gauge symmetry.
The new fermions $Q$ and $Q'$ are $SU(3)_c$ triplets and there are $p$ and $q$ flavors of them, respectively. 
With $p q_s + q q_t=0$, the theory is free from the $U(1)SU(3)_c^2$ gauge anomaly.
By rescaling the gauge coupling $g$, we have the freedom to set the charges to
\begin{equation}
q_s=q\ , \quad q_t=-p \ .
\end{equation}
The remaining $U(1)$ and $U(1)^3$ anomalies can be canceled away with additional color singlet fermions which have nothing to do with the physics to be described below.

Spontaneous symmetry breaking can be implemented using a scalar potential made of $|S|$ and $|T|$ which has an enhanced global symmetry $U(1)_S\times U(1)_T$.
The lowest-dimensional, gauge-invariant operator that violates the global symmetry (and in turn the PQ symmetry) is
\begin{equation}\label{eq:quality}
M_{\rm pl}^4 \left(S^{p} T^{q}/M_{\rm pl}^{p+q}\right)^{1/{\rm Gcd}(p,q)} \ ,
\end{equation}
where ${\rm Gcd}(p,q)$ is the greatest common divisor of integers $p$ and $q$.
For sufficiently large $p,q$ and small ${\rm Gcd}(p,q)$, the above operator is of high dimensions~\cite{Barr:1992qq}.
As a result, the $U(1)_S\times U(1)_T$ global symmetry is well protected and explicit PQ-breaking contributions to the axion potential is highly suppressed.
This is a truly appealing solution to the axion quality problem and many incarnations of the idea have been explored~\cite{Barr:1986hs, Fukuda:2017ylt, Duerr:2017amf, Ernst:2018bib, Fukuda:2018oco, DiLuzio:2020qio, Lu:2023ayc, Babu:2024udi, Babu:2024qzb}.

Below the PQ breaking scale, the scalar field vacuum expectation values (VEV) and angular excitations are
\begin{equation}\label{eq:VEV}
\begin{split}
S&= \frac{f_S}{\sqrt2} \exp\left[\frac{i}{f_S} \left(G\cos\theta - a\sin\theta \right)\right] \ , \\
T&= \frac{f_T}{\sqrt2} \exp\left[\frac{i}{f_T} \left(G\sin\theta + a\cos\theta \right)\right] \ ,
\end{split}
\end{equation}
where $\tan\theta = -pf_T/(qf_S)$. $G$ is the would-be Goldstone for the $Z'$ vector boson which has a mass $M_{Z'}=g\sqrt{q^2f_S^2+p^2 f_T^2}$, and $a$ is the physical axion field.

The quarks $Q, Q'$ acquire their masses from the Yukawa terms in Eq.~\eqref{eq:Yuk}. 
They contribute to the regular axion-gluon coupling, $(\alpha_3/8\pi)(a/f_a) G\widetilde G$, where the axion decay constant is
\begin{equation}
f_a = \frac{f_S f_T}{\sqrt{p^2 f_T^2 + q^2f_S^2}} \ .
\end{equation}
It is useful to note that the axion field space is not simply $2\pi f_a$, but rather $2\pi f_a \times {\rm min}(mp+nq)$, where $m,n$ are integers.
It can be found by uniformly shifting $a$ and $G$ fields in Eq.~\eqref{eq:VEV} that changes the phases of $S,T$ by $2\pi m$ and $2\pi n$, while leaving the VEVs invariant.
The theory thus has an axion domain wall number
\begin{equation}\label{eq:NDW}
N_{\rm DM} = {\rm min}(mp+nq) \ ,
\end{equation}
which counts the number of minima of the QCD axion potential.
If it is larger than 1, the theory suffers from the cosmological domain wall problem~\cite{Sikivie:1982qv}.
We will proceed by assuming that $p$ and $q$ are a coprime pair of integers, so that the axion domain wall number is 1 by virtue of the B\'ezout's identity~\cite{Bezouts}.
In other words, one can find special integers $m_*,n_*$ that set 
\begin{equation}\label{eq:domain1}
m_*p+n_*q=1 \ .
\end{equation}

An important prediction from the above theories in very early universe before the QCD phase transition is cosmic strings, cylindrical symmetric solutions of $S$ and $T$ fields. 
A generic string can be described with a pair of integers $(m,n)$, which corresponds to the phase factors of $S$ and $T$ varying by $2\pi m$ and $2\pi n$ as one goes around the center of string by a spatial angle $2\pi$.
The corresponding gauge field curls around with field value linear in $m$ and $n$~\cite{Hindmarsh:1994re, Vilenkin:2000jqa}.
There are two types of strings.
A pure gauge string $(m_g,n_g)$ corresponds to the special choice such that
\begin{equation}\label{eq:gaugestring}
m_gp+n_gq=0 \ .
\end{equation}
There is not a unique choice of $(m_g,n_g)$. 
Relevant for the following discussions is the lowest-energy gauge string with $(m_g, n_g) = (q, -p)$.
Alternatively, an axion string is characterized by an axion winding number,
\begin{equation}
w_a \equiv mp+nq = \text{nonzero integer} \ .
\end{equation}
For coprime $p,q$, there exists a string $(m_*,n_*)$ with unit axion winding number, $w_a=1$.
It is worth noting that even the theory has domain wall number $N_{\rm DM}=1$, it is still possible for the axion field to wind around a string by multiple times of $2\pi f_a$, resulting in $|w_a|>1$.

In the early universe, a networks of strings is produced along with the spontaneous symmetry breaking.
Without loss of generality, we assume $f_S>f_T$ hereafter and symmetry breaking occurs in two stages.
The $S$ field first obtains its VEV around time $t_S$ while the VEV of $T$ remains zero.
Only gauge strings are produced and the winding numbers $m$ may take all possible values. 
There is no axion field at this stage.
At a later time $t_T$, the second winding number $n$ emerges after the $T$ field gets a VEV.
The resulting string can be either gauge or axionic, depending on the $(m,n)$ values.
In particular, if a string is produced with axion winding number $|w_a|>1$, it will serve as the boundary of more than one domain walls once the QCD axion potential turns on~\footnote{In the literature, $w_a$ is sometimes referred to as the domain wall number of an axion string, in contrast to the domain wall number of the entire theory $N_{\rm DM}$, defined in Eq.~\eqref{eq:NDW}. Here we call $w_a$ the string's axion winding number for it not to be confused with $N_{\rm DM}$.}.
Based on the above picture, ref.~\cite{Barr:1992qq} further remarked (see also~\cite{Fukuda:2018oco}) that such theories still suffers from long-lived domain wall problem and cannot have a viable cosmology, even if $N_{\rm DW}$ is already set to unity.
This is a rather surprising statement because with $N_{\rm DM}=1$ the axion potential has a unique vacuum.
It is the job of the present work to point out a mechanism that allows the axion string-wall network to be destroyed.

Our proposal is based on observing a decomposition rule among different strings
\begin{equation}\label{eq:decomposition}
(m,n) = w_a (m_*, n_*) + c (m_g, n_g) \ ,
\end{equation}
which holds generally for any $(m,n)$.
Notably, the coefficient of the $(m_*, n_*)$ string must be equal to $w_a = mp+nq$, which can be validated using Eqs.~\eqref{eq:domain1} and \eqref{eq:gaugestring}.
The coefficient $c$ is needed for preserving the $S,T$ winding numbers.
With $(m_g,n_g)=(q,-p)$, $c$ is also an integer
\begin{equation}
c = m n_* - n m_* \ .
\end{equation}

Remarkably, Eq.~\eqref{eq:decomposition} implies that any string with axion winding number $|w_a|>1$ is topologically equivalent to $w_a$ unit-winding axion strings plus $c$ pure gauge strings. 
To deliver this point explicitly, we survey a number of concrete decomposition examples in table.~\ref{tab1}.
In all cases, at least one pure gauge string is needed for the decomposition to work, {\it i.e.}, $c\neq0$.
This property continues to hold for most strings with higher $(m,n)$ values not shown in the table.
We choose $p+q\geq10$ such that Eq.~\eqref{eq:quality} allows for a high quality axion, after taking into account of astrophysical constraints $f_S, f_T > f_a > 10^{9}\,$GeV~\cite{Adams:2022pbo}.

\begin{table}[h]
\centering
\begin{tabular}{||c|c|c|c|c|c||}
\hline\hline
 $p,q$ & $(m_*,n_*)$ & $(m_g,n_g)$ & $(m,n)$ & $w_a$ & $c$ \\
\hline
%p=3,q=7,m*=-2,n*=1,mg=7,ng=-3
& & & $(1,0)$ & 3 & 1 \\
& & & $(0,1)$ & 7 & 2 \\
%& & & $(-1,0)$ & $-3$ & $-1$ \\
%& $(0,-1)$ & $-7$ & $-2$ \\
\raisebox{1.2ex}[0pt]{$3,7$}& \raisebox{1.2ex}[0pt]{$(-2,1)$} & \raisebox{1.2ex}[0pt]{$(7,-3)$} & $(1,1)$ & 10 & 3 \\
& & & $(1,-1)$ & $-4$ & $-1$ \\
%\hline
%& & & $(1,0)$ & 5 & 1 \\
%& & & $(0,1)$ & 6 & 1 \\
%& & & $(-1,0)$& $-5$ & $-1$\\
%\raisebox{1.2ex}[0pt]{$5,6$} & \raisebox{1.2ex}[0pt]{$(-1,1)$} & \raisebox{1.2ex}[0pt]{$(6,-5)$} & $(0,-1)$ & $-6$ & $-1$ \\
%& & & $(1,1)$ & 11 & 2 \\
%& & & $(1,-1)$ & $-1$ & 0 \\
\hline
& & & $(1,0)$ & 5 & 3 \\
& & & $(0,1)$ & 7 & 4 \\
%& & & $(-1,0)$& $-5$ & $-3$\\
%& $(0,-1)$ & $-7$ & $-4$ \\
\raisebox{1.2ex}[0pt]{$5,7$} & \raisebox{1.2ex}[0pt]{$(-4,3)$} & \raisebox{1.2ex}[0pt]{$(7,-5)$} & $(1,1)$ & 12 & 7 \\
& & & $(1,-1)$ & $-2$ & $-1$ \\
\hline
& & & $(1,0)$ & 9 & $-2$ \\
& & & $(0,1)$ & 4 & $-1$ \\
%& & & $(-1,0)$& $-9$ & 2\\
%& $(0,-1)$ & $-4$ & 1 \\
\raisebox{1.2ex}[0pt]{$9,4$} & \raisebox{1.2ex}[0pt]{$(1,-2)$} & \raisebox{1.2ex}[0pt]{$(4,-9)$} & $(1,1)$ & 13 & $-3$ \\
& & & $(1,-1)$ & $5$ & $-1$ \\
%\hline
%& & & $(1,0)$ & 1 & $0$ \\
%& & & $(0,1)$ & 12 & $-1$ \\
%%& & & $(-1,0)$& $-1$ & 0\\
%%& $(0,-1)$ & $-12$ & 1 \\
%\raisebox{1.2ex}[0pt]{$1,12$} & \raisebox{1.2ex}[0pt]{$(1,0)$} & \raisebox{1.2ex}[0pt]{$(12,-1)$} & $(1,1)$ & 13 & $-1$ \\
%& & & $(1,-1)$ & $-11$ & $1$ \\
\hline\hline
\end{tabular}
\caption{\it Examples of string decomposition with the lowest $(m,n)$ values that are most likely produced in the early universe. 
Negative $w_a$ and/or $c$ means decomposing into conjugate strings with opposite winding numbers.}
\label{tab1}
\end{table}

The immediate next physical question is whether the disintegration of $|w_a|>1$ strings into the $(m_*, n_*)$ and $(m_g, n_g)$ ones can take place sufficiently early, without conflicting with cosmological observations.
If the answer is positive, each resulting $w_a=1$ axion string will be the edge of a single domain wall and they can be destroyed as usual~\cite{Sikivie:2006ni}, leaving no long-lived domain walls.

There are two possible ways for the above string decomposition to occur.
The first is spontaneous decay.
To assess the feasibility, we resort to an analytic estimate of string tension (energy per unit length) given in~\cite{Niu:2023khv}.
In the $f_S\gg f_T$ limit, the string tension is approximately $\mu \sim 2 \pi m^2 f_S^2$.
Applying it to both sides of Eq.~\eqref{eq:decomposition}, we find there is an energy barrier preventing the decay if $m^2 < |w_a| m_*^2 + |c| m_g^2$ which is likely the case because strings with the lowest $m,n$ values are the most energetically favored to populate around the symmetry breaking temperatures.
Indeed, a barrier exists for all examples listed in table.~\ref{tab1}.

The second way of string decomposition involves the axion domain walls which come to life when the QCD axion potential turns on.
With domain wall number $N_{\rm DM}=1$, the axion field value changes by $\pm 2\pi f_a$ across a wall. 
Hence, a $|w_a|>1$ string first appears as the boundary of $|w_a|$ domain walls.
Crucially, these domain walls, instead of balancing with each other, will do the job of string decomposition as dictated by Eq.~\eqref{eq:decomposition}.
Domain walls have energy per unit area $\sigma \sim m_a f_a^2$.
The time scale when the domain wall energies exceed the energy barrier of string decomposition can be estimated as
%
%\begin{equation}\label{eq:t2}
%\frac{t_2}{\rm sec} \sim \left\{ \begin{array}{cc}
%\displaystyle10^{-6} \left(\frac{f_T}{10^{12}\,\rm GeV}\right)^{\frac{1}{3}}\left(\frac{f_S}{f_T}\right)^{\frac{2}{3}}, & f_T \ll f_S \lesssim 300 f_T \\
%\displaystyle10^{-10} \left(\frac{f_T}{10^{12}\,\rm GeV}\right)\left(\frac{f_S}{f_T}\right)^2, & f_S \gtrsim 300 f_T
%\end{array} \right.
%\end{equation}
%
\begin{equation}\label{eq:t2}
\frac{t_2}{\rm sec} \sim \left\{ \begin{array}{cc}
\displaystyle10^{-6} f_{T,12}^{1/3}\left(\frac{f_S}{f_T}\right)^{2/3}, & 1 \ll f_S/f_T \lesssim 300f_{T,12}^{-1/2} \\
\displaystyle10^{-10} f_{T,12} \left(\frac{f_S}{f_T}\right)^2, & f_S/f_T \gtrsim 300f_{T,12}^{-1/2}
\end{array} \right.
\end{equation}
where $f_{T,12}\equiv{f_T}/({10^{12}\,\rm GeV})$, we use time dependent axion mass~\cite{Sikivie:2006ni}, and $f_a\sim f_T$ for $f_T\ll f_S$.
After $t_2$, the domain walls carry sufficient energy for the above string decomposition to occur.
Each resulting $w_a=\pm1$ axion string is connected to one domain wall so that they quickly decay away.
A schematic picture is shown in Fig.~\ref{fig:picture}.
Compared to the standard scenario~\cite{Sikivie:2006ni}, here $t_2$ is enhanced by a factor of $(f_S/f_T)^2$.
However, there is still ample parameter space with $f_S<10^{5}f_T$ that allows the string decomposition and string-wall decay to complete well before the big-bang nucleosynthesis (BBN).

In the above discussion of depleting high $|w_a|$ strings to avoid the domain wall problem, a key player is the need for pure gauge strings.
There are two epochs when the gauge strings prevail throughout the universe.
One happens early on $t_S < t < t_T$, where only the $S$ field develops a VEV thus only gauge strings are present.
The second epoch corresponds to $t> t_2$ scale where the axion domain walls decompose the strings following the rule of Eq.~\eqref{eq:decomposition}.
The decomposition liberates gauge strings. 
While all the resulting axion strings are quickly destroyed, gauge strings do not talk to axion domain walls and survive until later times.
On the other hand, during $t_T< t< t_2$, almost all strings filling the universe are axion strings with $w_a\neq0$ (see {\it e.g.} table.~\ref{tab1}).

\begin{figure}[t]
  \includegraphics[width=1\columnwidth]{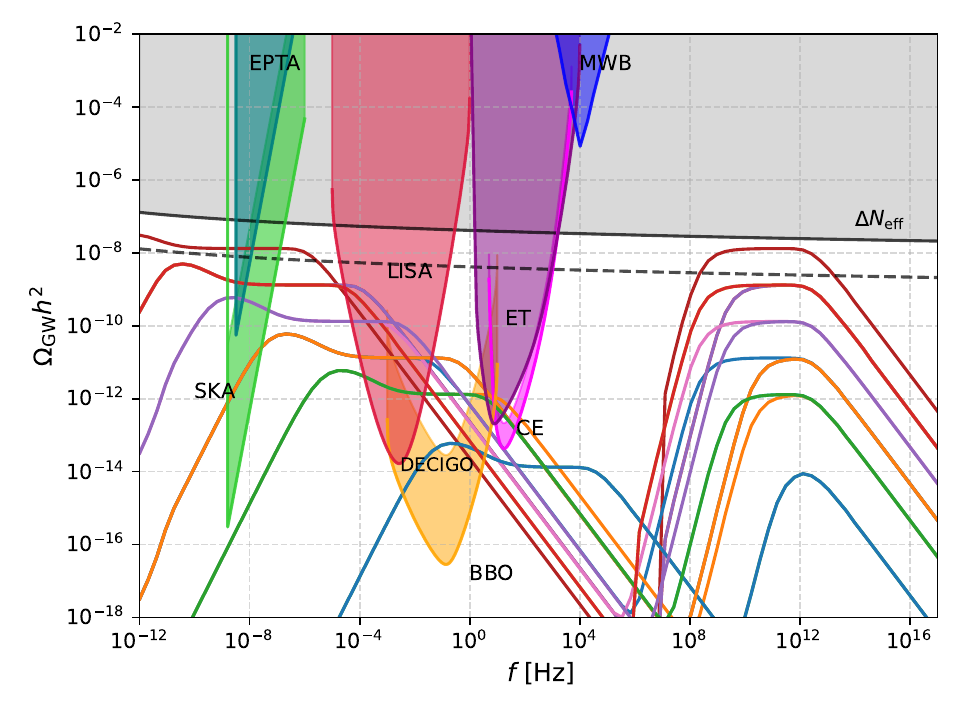}
  \caption{Colourful curves show gravitational waves spectrum from gauge string radiation with two plateaus as a generic prediction of the Barr-Seckel models for high-quality axion.
   We scan the two decay constants $f_S, f_T$ in the window between $10^9$ and $10^{16}$\,GeV, while keeping $f_T<f_S<10^5 f_T$ so that all axion strings and domain walls are destroyed before BBN (see Eq.~\eqref{eq:t2}).
   We show the sensitivity curves of future GW interferometers, pulsar time arrays, magnetic Weber bar, as well as cosmological probes of extra radiation ($\Delta N_{\rm eff}$).
   The grey region is excluded by the existing limit $\Delta N_{\rm eff}\lesssim0.3$~\cite{Planck:2018vyg}. 
  } \label{fig:GW}
\end{figure}

A crucial phenomenological distinction between pure-gauge and axion strings lies in that, gauge strings radiate GWs whereas the radiation of axion strings is mainly in the form of axion particles~\cite{Davis:1986xc, Vilenkin:1986ku, Hagmann:1990mj, Preskill:1992ck, Battye:1993jv, Yamaguchi:1998gx, Hagmann:2000ja, Gorghetto:2018myk}.
As a result, we will show that the proposed mechanism leads to a characteristic GW spectrum. 
After birth, the network of strings quickly evolve to the scaling solution, in the spirit of the Kibble mechanism~\cite{Kibble:1976sj}.
Within a Hubble patch, interactions of strings create loops that eventually evaporate away.
We adopt a semi-analytic approach presented in~\cite{Cui:2018rwi, Cui:2017ufi} to model the gauge string loop production and decay.
The GW spectrum as a function of frequency $f$ is defined as $\Omega_{\rm GW}(f) \equiv d\rho_{\rm GW}/d \ln f/\rho_c$~\cite{Mingarelli:2019mvk}, where $\rho_{\rm GW}$ is the GW energy density and $\rho_c$ is the critical density of the universe.
The contribution from gauge string radiation is
\begin{widetext}
\begin{equation}\label{eq:GWspectrum}
\Omega_{\rm GW} = \sum_{k=1}^{\infty} \frac{1}{\rho_c}\frac{2k}{f} \frac{F_\alpha \Gamma^{(k)} G \mu^2}{\alpha(\alpha + \Gamma^{(k)} G \mu)} \int_{t_S}^{t_0} dt \frac{C_{eff}}{t_i^4} \left(\frac{a(t)}{a(t_0)}\right)^5 \left(\frac{a(t_i)}{a(t)}\right)^3 
\Theta(t-t_i) \Theta(t_i-t_S) \Theta\left(t_i - \frac{\Gamma G \mu t}{\alpha + \Gamma G \mu} \right) W(t_i) \ ,
\end{equation}
\end{widetext}
where $G$ is the Newton's constant, $\alpha \simeq F_\alpha \simeq 0.1$, $C_{\rm eff}\simeq 5.4$, $\Gamma\simeq 50$, $\Gamma^{(k)} =\Gamma k^{-4/3}/\zeta(4/3)$, and the string tension $\mu$ is of order $f_S^2$.
For given GW frequency $f$, gauge string loops produced at time 
\begin{equation}\label{eq:ti}
t_i = \frac{1}{\alpha + \Gamma G \mu}\left[\Gamma G \mu t + \frac{2k}{f} \frac{a(t)}{a(t_0)}\right] \ ,
\end{equation}
decay at time $t$.
The latter is integrated from the birth of the first gauge strings ($t_S$) till today ($t_0$). 
$a$ is the scale factor of the universe. $\Theta$ is a unit step function.
To apply it to the theories considered here, we further insert a window function $W(t_i)$ to model the two epochs of gauge strings production in the early universe
\begin{equation}
W(t_i) = \Theta\left(t_i -t_S\right)\Theta\left(t_T-t_i\right) + \Theta\left(t_i-t_2\right) \ ,
\end{equation}
where $t_2$ is the axion string-wall collapsing time and given by Eq.~\eqref{eq:t2}.
Throughout the calculation, we assume the universe is radiation dominated at early times. At time $t_{S,T}$, the universe has a temperature $f_{S,T}$, respectively.

The resulting GW spectrum is shown in Fig.~\ref{fig:GW}.
It features two plateaus and a valley in between, which can be understood as the following.
Using Eq.~\eqref{eq:GWspectrum}, one can show that gauge string loops produced around a time $t_i$ make a triangle shaped GW spectrum in the log-log plot of $\Omega_{\rm GW}$ versus $f$, peaked at $f_* \simeq 10^{13} \,{\rm Hz} \times [T(t_i)/f_S]$, where $T(t_i)$ denotes the temperature of the universe at time $t_i$.
It corresponds to the two terms in the square bracket of Eq.~\eqref{eq:ti} are comparable.
In the $\alpha \gg \Gamma G f_S^2$ limit, GW is dominantly produced with original frequency $(\alpha t_i)^{-1}$ around time $t\simeq \alpha t_i/(\Gamma G f_S^2)$.
Thus a larger $f_S$ brings $t$ closer to $t_i$, resulting in more redshift till today and smaller $f_*$.
The spectrum grows as $f^3$ for $f<f_*$ and falls as $1/f$ for $f>f_*$.
For string loop production lasting over an extended period, the corresponding triangles stack next to each other to form a plateau.
The two stages of gauge string production discussed above leads to two plateaus.
With $T(t_2) \leq T(t_{\rm QCD})\simeq 0.1\,$GeV, the lower-$f$ plateau extends up to a frequency $f \lesssim 1 \,{\rm Hz} \times (10^{12}\,{\rm GeV}/f_S)$. 
The higher-$f$ plateau spans a window $10^{13} \,{\rm Hz} \times (f_T/f_S) \lesssim f \lesssim 10^{13} \,{\rm Hz}$.
For frequencies in between, the GW spectrum is suppressed due to the lack of gauge strings during $t_T<t_i<t_2$.
With the astrophysical bound $f_T\sim f_a > 10^9\,$GeV, the valley always exists.
Therefore, the shape of $\Omega_{\rm GW}h^2$ shown in Fig.~\ref{fig:GW} resembles a robust prediction of the Barr-Seckel models with any coprime $(p,q)$ pairs.

This prediction can be tested by a vast landscape of upcoming GW experiments depicted in Fig.~\ref{fig:GW}~\cite{Schmitz:2020syl}.
The lower-$f$ plateau is the target of future GW interferometers including LISA~\cite{LISA:2017pwj}, Big Bang
Observer (BBO)~\cite{Corbin:2005ny}, DECIGO~\cite{Kawamura:2020pcg}, Einstein Telescope (ET)~\cite{Abac:2025saz}, Cosmic Explorer (CE)~\cite{Reitze:2019iox}, as well as pulsar timing arrays including SKA~\cite{Janssen:2014dka} and EPTA~\cite{EPTA:2015qep}.
The interferometers experiments can also observe the downdfall of $\Omega_{\rm GW}h^2$ towards the valley.
The height of plateaus is proportional to the larger decay constant $f_S$.
For $f_S \gtrsim 10^{16}$\,GeV, the higher-$f$ plateau contributes to extra radiation ($\Delta N_{\rm eff}$)~\cite{Maggiore:2000gv, Smith:2006nka, Blas:2021mqw} and can be targeted by future cosmic microwave background observations with the Simons Observatory~\cite{SimonsObservatory:2025wwn} and CMB-S4~\cite{CMB-S4:2016ple}.
It is also interesting to see proposals to probe high frequency GW with novel detectors~\cite{Aggarwal:2025noe}.
A joint effort of these experiments will be crucial to uncover the unique string cosmology as consequence of the Barr-Seckel models for high-quality axion.

In summary, we point out a string decomposition mechanism that eliminates long-lived domain walls in high-quality axion models where a gauged $U(1)$ symmetry is spontaneously broken by two scalar fields with distinct charges.
It renders viable cosmology to a broader range of theories with various numbers of heavy quarks. 
% and can be tested by future GW experiments.
%
Our result provides useful guide for further model building, cosmological simulations, and experimental searches.
We end with a comment on the case of axion serving as dark matter filling the universe.
If the axion string-wall collapsing time $t_2$ [defined in Eq.~\eqref{eq:t2}] lies close to $t_{\rm QCD}$, the axion particle production from strings is similar to recent calculations by~\cite{Buschmann:2021sdq, Benabou:2024msj}, with the favored axion mass around tens to hundreds of $\mu$eV.
For higher ratios $f_S/f_T \gtrsim 10^3$ and in turn a larger $t_2$, the decay of domain walls could be the dominant contribution to the axion relic which points toward a heavier axion dark matter and requires in-depth simulation.

{\it Acknowledgement --} We thank Kaladi Babu, Malte Buschmann, Roni Harnik, and Miha Nemev\v{s}ek for useful discussions.
This work is supported by a Subatomic Physics Discovery Grant (individual) from the Natural Sciences and Engineering Research Council of Canada.
This work was performed in part at the Aspen Center for Physics, which is supported by National Science Foundation grant PHY-2210452.

\bibliographystyle{apsrev4-1}
\bibliography{ref}{}
\end{document}